 \documentclass[journal]{IEEEtran}
\ifCLASSINFOpdf
\else
\fi

\usepackage{xcolor}

\usepackage{booktabs}
\usepackage{array}
\usepackage{multirow}
\usepackage{adjustbox}
\begin{document}
%
\title{Feature-Centric Approaches to Android Malware Analysis: A Survey}


\author{Shama Maganur,			 
        Yili~Jiang,~\IEEEmembership{Member,~IEEE,}
			Jiaqi~Huang,~\IEEEmembership{Member,~IEEE,}
      
  and~Fangtian~Zhong,~\IEEEmembership{Member,~IEEE}
			\thanks{Shama Maganur is with the Gianforte School of Computing,
				 Montana State University, Bozeman,
				MT 59717, USA. E-mail: shama.maganur@gmail.com}
			\thanks{Yili~Jiang is with the Department of Computer Science, Georgia State University, Atlanta, GA 30303, USA. E-mail: yjiang27@gsu.edu}
                \thanks{Jiaqi~Huang is with the Department of Computer Science and Cybersecurity, University of Central Missouri, Warrensburg, MO 64093, USA. E-mail: jhuang@ucmo.edu}
            \thanks{Fangtian~Zhong (Corresponding author) is with the Gianforte School of Computing,
				 Montana State University, Bozeman,
				MT 59717, USA. E-mail: fangtian.zhong@montana.edu}
        }

%


\maketitle

\begin{abstract}
Sophisticated malware families exploit the openness of the Android platform to infiltrate IoT networks, enabling large-scale disruption, data exfiltration, and denial-of-service attacks. This systematic literature review (SLR) examines cutting-edge approaches to Android malware analysis with direct implications for securing IoT infrastructures. We analyze feature extraction techniques across static, dynamic, hybrid, and graph-based methods, highlighting their trade-offs: static analysis offers efficiency but is easily evaded through obfuscation; dynamic analysis provides stronger resistance to evasive behaviors but incurs high computational costs, often unsuitable for lightweight IoT devices; hybrid approaches balance accuracy with resource considerations; and graph-based methods deliver superior semantic modeling and adversarial robustness. This survey contributes a structured comparison of existing methods, exposes research gaps, and outlines a roadmap for future directions to enhance scalability, adaptability, and long-term security in IoT-driven Android malware detection.
\end{abstract}


%
\IEEEpeerreviewmaketitle

\section{Introduction}
The proliferation of Android devices, commanding over 70\% of the global mobile operating system market share as of 2018 \cite{paper56}, has made them a prime target for malware, with approximately 12,000 new samples detected daily \cite{paper56}. The open-source nature of Android has led to an exponential increase in sophisticated malware, including ransomware, spyware, and botnets, posing significant threats to user privacy, data security, and system integrity \cite{paper45,paper51,paper53}. Traditional malware detection methods, such as signature-based and heuristic-based approaches, struggle to keep pace with the volume, velocity, and variety of modern malware, particularly zero-day attacks and adversarial evasions \cite{paper27,paper43,paper60}. Consequently, machine learning (ML) techniques have emerged as powerful tools, leveraging static, dynamic, hybrid and graph-based feature extraction methods to enhance detection accuracy and robustness \cite{paper1, paper4, paper13, paper48, paper52}. Static analysis techniques, such as permission analysis, API call extraction, and bytecode-to-image conversion, offer efficient and scalable solutions but are vulnerable to obfuscation and adversarial manipulations \cite{paper5, paper16, paper20, paper27, paper43}. Dynamic analysis, capturing runtime behaviors like system calls and network traffic, provides resilience against code obfuscation but faces challenges in code coverage and resource intensity, making it less suitable for resource-constrained IoT Android devices \cite{paper4, paper25, paper38, paper63}. Hybrid approaches combine the strengths of static and dynamic methods to improve detection accuracy and adaptability, as seen in frameworks like Hybroid and C2Miner \cite{paper13, paper61}. Meanwhile, graph-based learning, utilizing structures like function call graphs (FCGs) and heterogeneous graphs, captures complex semantic relationships, offering robustness against adversarial attacks and obfuscation \cite{paper48, paper50, paper52, paper53}. However, each method presents unique trade-offs in terms of computational efficiency, scalability, and resistance to evolving threats, necessitating a comprehensive evaluation of their feature extraction strategies.

The rapid evolution of Android malware underscores the need for a systematic review of feature extraction techniques across these analysis paradigms. Existing surveys, such as \cite{paper56}, provide valuable overviews of ML-based detection frameworks but often lack a focused comparison of feature-specific strengths and limitations. This survey addresses this gap by analyzing 68 carefully selected papers from 2009 to 2025, sourced from IEEE Xplore, ACM Digital Library, and SpringerLink, focusing on static, dynamic, hybrid, and graph-based feature extraction methods for Android malware detection. By synthesizing insights from these works, we elucidate the effectiveness, challenges, and opportunities of diverse feature extraction techniques, ultimately providing a roadmap for researchers and practitioners to develop robust, efficient, and scalable malware detection systems for Android ecosystems.

\section{Related Work}

\subsection{Differences with Previous Related Surveys}
Prior surveys, such as \cite{paper56}, have comprehensively reviewed Android malware detection techniques, emphasizing ML and DL frameworks and their general applicability. However, they often focus on broad methodologies without delving into the specific contributions of feature extraction techniques across static, dynamic, hybrid, and graph-based paradigms. For instance, \cite{paper56} outlines static, dynamic, and hybrid analysis but does not provide a detailed comparison of feature-specific performance in resource-constrained Android settings. Similarly, other surveys lack an in-depth analysis of adversarial robustness and the impact of feature selection on detection efficacy \cite{paper43, paper44, paper55}. This survey addresses these gaps by:
\begin{itemize}
\item Providing a detailed evaluation of feature extraction techniques across four analysis paradigms, with a focus on their applicability to Android devices.
\item Analyzing the robustness of these techniques against adversarial attacks, such as those explored in \cite{paper43, paper44, paper53, paper55}, which are critical in dynamic threat landscapes.
\item Incorporating recent advancements (up to 2025) to address emerging challenges like ransomware and collusion attacks \cite{paper45, paper54}.
\item Offering a structured comparison of feature extraction methods based on accuracy, computational efficiency, and scalability, derived from the 68 reviewed papers.
\end{itemize}
As a result, this work not only bridges the identified shortcomings but also delivers a comprehensive roadmap, enabling researchers and practitioners to develop more resilient and efficient Android malware detection systems.

\section{BACKGROUND}
\subsection{Program Representation}
Android malware analysis requires representing APK files at various abstraction levels to facilitate feature extraction for detection and classification, as these representations capture essential structural, behavioral, or semantic aspects of the app across static, dynamic, hybrid, and graph-based techniques. However, a key challenge lies in balancing abstraction to handle complexities like obfuscation while maintaining efficiency and accuracy in resource-constrained environments. To address this, three primary representation types are employed: bytecode for raw components, intermediate representation for transformed analyzable formats, and source code for decompiled content, each enabling specific analysis methods as illustrated in the reviewed papers. Ultimately, this layered approach resolves representation gaps by providing flexible, obfuscation-resistant foundations that enhance detection robustness and scalability in Android ecosystems.
\subsubsection{bytecode}
Binary representations involve raw APK components without requiring decompilation or execution. Examples include DEX bytecode analyzed for entropy, asset files, native binaries, runtime system call traces, network traffic flows or PCAP files, and manifest permissions/intents. Feature representations can also take the form of grayscale or Markov images derived from bytecode/opcodes, opcode sequences, n-grams, entropy vectors,  histograms, permission vectors, or system call abstractions.
\subsubsection{intermediate representation}
Intermediate representations transform binaries into program representation graphs for analysis, such as call graphs, system dependence graphs or heterogeneous graphs.
\subsubsection{source code}
Source code representations involve decompiled content, such as Smali or Java code from APKs, used for extracting APIs, commands, or patterns. However, direct source code analysis is less common, often supplemented by intermediate forms to handle obfuscation.

\subsection{Static Analysis Techniques}
Static analysis examines Android app artifacts without execution, extracting features like permissions, APIs, opcodes, and structural patterns from APK files. These techniques are prevalent in the reviewed papers for efficient malware detection, though they are vulnerable to obfuscation.

\subsubsection{flow analysis}  
It examines the propagation of data and control within a program to identify potential issues or optimization opportunities. Examples include API call flows \cite{paper39}, where dominance trees are used to mine flow patterns. This type of analysis can reveal suspicious dependencies, such as sensitive API invocations that may indicate malicious intent, thereby enabling early optimization of detection models by pinpointing control paths vulnerable to exploitation. However, its effectiveness depends on accurate program modeling and may overlook runtime variations that are not visible in static code.

\subsubsection{call-graph analysis}
Call-graph analysis models the calling relationships between functions or subroutines in a program as a graph, where nodes are procedures and edges represent calls, aiding in understanding control flow and interprocedural behavior. It is common in the papers, such as Function Call Graphs (FCG) in \cite{paper30}(structural features), \cite{paper8} (Sensitive Function Call Graphs with GCN), \cite{paper36} (GCN on call graphs), \cite{paper39} (dominance trees from call graphs via Soot/FlowDroid), and \cite{paper50} (FCG with opcode vectors).

\subsubsection{pattern match}
Pattern matching checks sequences or structures for predefined patterns, such as signatures or behavioral motifs, to identify matches indicative of malware. The papers include semantic patterns in \cite{paper23} (API call chain patterns), hyperlink/similarity relations in \cite{paper14} (pattern match for API selection), and code patterns in \cite{paper11} (forensic packaging patterns).

\subsubsection{Software composition analysis (SCA)} Software composition analysis (SCA) identifies and analyzes open-source or third-party components in software to detect vulnerabilities, licenses, or risks. In Android malware analysis, it examines app dependencies and libraries for potential security issues without execution.

\subsection{Dynamic Analysis Techniques}
Dynamic analysis executes Android apps in controlled environments, such as sandboxes or emulators, to observe runtime behaviors like system calls, network activity, and resource usage. These methods complement static approaches by capturing evasive actions that manifest only during execution, though they incur higher overhead and risk of detection evasion.

\subsubsection{System Call Tracing}
System call tracing monitors interactions between an app and the operating system kernel during execution to capture low-level behaviors, such as file access or process creation \cite{enck2014taintdroid}. In the papers, it is used for dynamic feature extraction, e.g., ioctl calls in \cite{paper4}, runtime patterns in \cite{paper13}, abstractions in \cite{paper25}, and sub-traces in \cite{paper49}.
\subsubsection{Network Traffic Analysis}
Network traffic analysis examines packet flows, HTTP requests, or communication patterns to detect malicious C2 servers or data exfiltration \cite{arp2014drebin}. Examples include encrypted traffic in \cite{paper12} (POPNet), flow statistics in \cite{paper22}, prioritized features in \cite{paper38}, HTTP in \cite{paper58}, and OSR on HTTP in \cite{paper67, paper68}.
\subsubsection{Behavioral Profiling}
Behavioral profiling logs high-level actions like UI interactions, API calls, and permission usage during runtime to model app behavior \cite{yan2018survey}. It is applied in \cite{paper41} for user-oriented traces and in hybrids like \cite{paper13}.
\subsubsection{Dynamic Symbolic Execution}
Dynamic symbolic execution, also known as concolic testing \cite{sen2005cute}, is a hybrid approach where the program is executed with concrete inputs while simultaneously performing symbolic execution to explore additional paths.

\subsection{Graph Learning Representation Techniques}
Graph learning represents Android apps as graphs (e.g., nodes as APIs/functions, edges as calls) to capture relational features, using neural networks like GNNs for embedding and classification. These methods excel in modeling complex structures but require careful graph construction to handle heterogeneity and scale. 
\subsubsection{Graph Convolutional Network (GCN)}
Graph convolutional networks extend convolutions to graphs by aggregating neighborhood features via spectral or spatial methods \cite{kipf2017semi}. 
\subsubsection{Heterogeneous Graph Neural Network (HetGNN)}
Heterogeneous graph neural networks handle graphs with multiple node/edge types by type-specific aggregations or attentions \cite{zhang2019heterogeneous}. 
\subsubsection{Variational Graph Autoencoder (VGAE)}
Variational graph autoencoders extend VAEs to graphs for unsupervised embedding, learning latent representations via reconstruction and KL divergence \cite{kipf2016variational}.

\section{A SUMMARY OF REVIEWED PAPERS}
\subsection{Selection Rules}
We surveyed publications from 2009 to 2025 using the targeted keywords ``Android malware detection” and ``Android malware classification” across major academic databases, including IEEE Xplore, ACM Digital Library, and others, yielding a total of 1,664 papers. As an initial step, we applied a length-based filter, retaining only papers with at least 10 pages to ensure sufficient depth and substance. This reduced the set to 96 candidates. We then applied a relevance filter through a detailed review of abstracts. From this stage, 68 papers were identified as truly relevant, while 4 duplicates, 6 focused on non-Android malware, and 19 unrelated studies were excluded.

\subsection{Classification Criteria}
To systematically evaluate the 68 selected papers on Android malware detection and classification, we developed a multi-dimensional classification scheme that organizes them by core analysis technique (static, dynamic, hybrid, or graph-based), underlying methodology, performance metrics, and adoption frequency. This structured approach reveals clear patterns: static techniques, exemplified by permission-based machine learning and opcode analysis, stand out for their lightweight design and widespread use in on-device deployment, delivering efficient results as demonstrated in \cite{paper5, paper6, paper24}, though they are prone to obfuscation challenges. Dynamic methods, such as system call tracing, excel in uncovering runtime behaviors that expose hidden threats, yet they introduce computational overhead, as evidenced in \cite{paper4, paper13, paper25}. Hybrid approaches bridge these worlds by fusing static efficiency with dynamic depth, often achieving superior accuracy through multi-view integration \cite{paper13, paper41, paper42}. Graph learning techniques, meanwhile, construct relational models of app structures to enhance obfuscation resistance, with graph convolutional network (GCN) embeddings showing strong potential in adversarial contexts \cite{paper8, paper36, paper52, paper51}. Performance metrics across the papers consistently highlight accuracy levels exceeding 95\% in leading models, alongside F1-scores for handling imbalances, while deep learning hybrids like CNN+DNN are increasingly adopted for real-time applications \cite{paper21}. This framework not only organizes the literature into a coherent narrative but also pinpoints critical gaps, such as static methods' limitations against zero-day threats, thereby guiding future efforts toward more adaptive and explainable systems. The detailed classification is summarized in the table below, offering an evidence-based synthesis of methodologies and metrics from the reviewed summaries.

\subsection{Benchmark Datasets}
Datasets supply the essential ground truth, spanning malware families, benign apps, and threat evolutions and enabling models to train and test effectively. Early research leaned on foundational sets like Drebin and Genome for static validations, while later works shifted to modern collections such as CICMalDroid and AndroZoo to tackle zero-day and obfuscated variants. This transition mirrors the field's progression toward managing dynamic, large-scale data, yet it exposes persistent hurdles like class imbalances, outdated samples, and the demand for cross-dataset testing.

Drebin stands as the most utilized benchmark, appearing in 39 papers for frameworks blending permissions, APIs, and intents \cite{arp2014drebin}. With 5,560 malware samples across 179 families plus Google Play benign apps (2010–2012), it facilitates explainable ML. VirusShare follows in 20 papers for dynamic and graph analyses, offering over 24,000 samples to highlight real-time threats. AndroZoo supports 14 papers with time-stamped, VirusTotal-verified APKs \cite{AndroZooCollecting}, ideal for app evolution studies.

CICMalDroid aids 7 papers in hybrid/pruning tests, featuring 17,000+ samples (2018+) across adware/banking/SMS/riskware  \cite{paper5}. Genome/MalGenome appears in 12 for behavioral family classification, with 1,260 samples from 49 families\cite{arp2014drebin}. Other datasets include CICAndMal2017 for ransomware traffic (402,834 records), AMD for families (24,553 samples)\cite{SMMARWAR2024100130}, and specialized ones like RansomProber for OCR \cite{paper45} or custom crawls for evasion \cite{paper2}.

This dataset diversity narrates a maturation arc: from Drebin's baselines to CICMalDroid's fragmentation handling \cite{paper19}. Yet, imbalances prompt oversampling \cite{paper22}, outdated data drives augmentation \cite{paper11}, and cross-evaluations show 5–10\% drops \cite{paper8,paper20}. Moving forward, prioritize temporal metadata to close emerging threat gaps, bolstering techniques like weighted graphs \cite{paper52} against polymorphic malware.

\section{Malware Analysis Across Features}
There are four primary features used to detect malware: static analysis, dynamic analysis, hybrid analysis, and graph representation learning. These methods collectively enhance the detection of malware by addressing different aspects and potential weak points in software security.

\subsection{Static Analysis Features} 

\begin{figure}[htbp]
\centering
\includegraphics[width=\columnwidth]{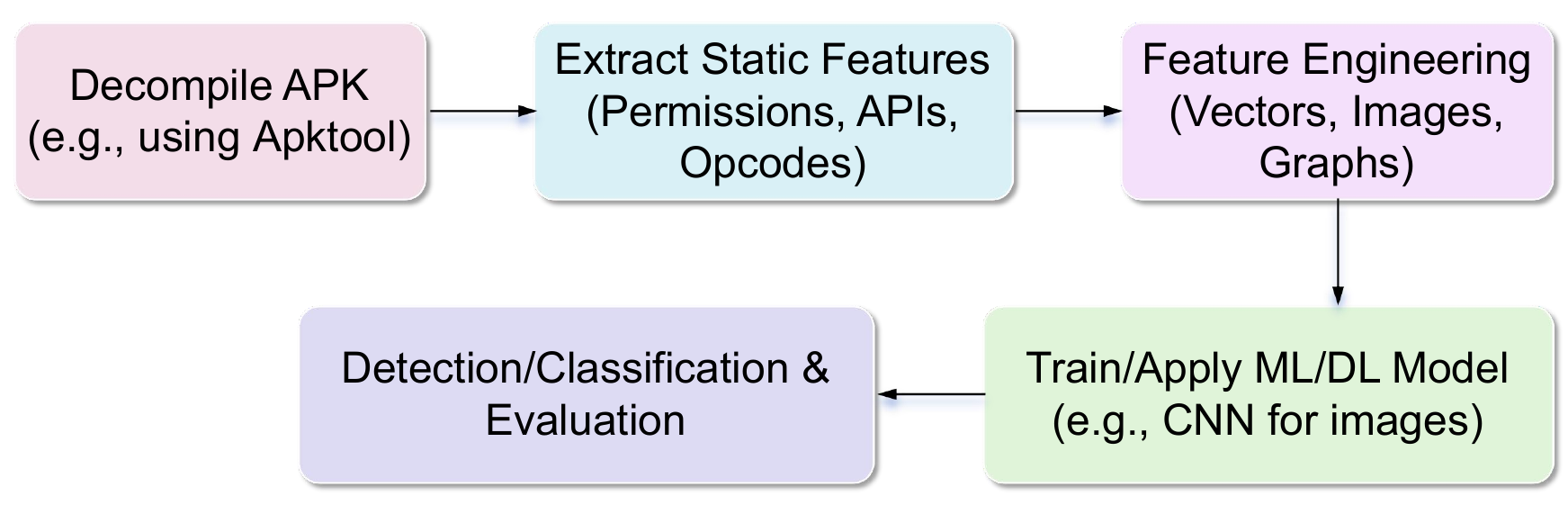}
\caption{Workflow for Static Analysis Technique in Android Malware Detection.}
\label{fig:static-workflow}
\end{figure}

Static analysis provides an execution-free workflow for Android malware detection, offering efficiency and scalability that is particularly important for resource-constrained environments such as IoT devices. As shown in Fig. \ref{fig:static-workflow}, the process begins with APK decompilation using tools like Apktool or Androguard, which grants access to Smali code, DEX bytecode, manifest files, and binaries without exposing the system to runtime vulnerabilities \cite{paper1,paper2,paper3,paper11,paper20,paper21,paper30,paper36}.  Feature extraction follows, leveraging a variety of representations: permissions are vectorized for lightweight analysis \cite{paper5,paper47}; opcodes and n-grams are sequenced to capture structural patterns \cite{paper3,paper57}; entropy histograms detect obfuscation \cite{paper17}; and APIs or function call graphs (FCGs) encode semantic behavior \cite{paper8,paper30,paper36}. Feature engineering refines these into inputs for machine learning or deep learning models, including grayscale or Markov images for CNN-based analysis \cite{paper1,paper20,paper21,paper64}, vectors for traditional ML, and pruned graph representations to reduce dimensionality \cite{paper62,paper66}. Models include DNNs for permissions \cite{paper40}, Random Forests for opcode sequences, and Transformer-based networks for semantic patterns \cite{paper37}. 

These static analysis principles are adapted in IoT-specific frameworks to balance precision with efficiency. For example, ViT4Mal \cite{10.1145/3609112} converts bytecode into images and applies a lightweight Vision Transformer (ViT) to detect malware efficiently on edge devices. Hardware optimizations like quantization, loop pipelining, and array partitioning enhance efficiency without sacrificing precision. The system preprocesses bytecode into image patches, passes them through transformer encoder blocks (using layer normalization, multi-head attention, skip connections, and MLP layers), and classifies binaries as benign or malicious via a decoder network. Similarly, TransMalDE integrates IoT devices, edge nodes, and cloud servers within the static analysis workflow. IoT devices perform disassembly and lightweight feature extraction, identifying sensitive API calls and constructing FCGs to capture malware behavior. Edge servers monitor inputs in real time, transform sensitive APIs into semantic vectors via TF-IDF, and classify suspicious behaviors using a Transformer-based detection model. Upon detecting malware, the edge server issues user alerts and executes automated control actions on the local infrastructure. The cloud aggregates data from edge nodes to train large-scale Transformer models, maintain malware databases, and manage resources during attacks. Its multi-head attention captures semantic relationships in subgraph features, while outputs are merged through a linear layer with Softmax for final classification.

\subsubsection{Comparison Among Different Static Analysis Features}
Static analysis techniques for Android malware detection employ a wide variety of feature representations and learning models, each with specific strengths and trade-offs (see Table~\ref{tab:feature-comparison}). Image-based approaches convert bytecode or opcode sequences into visual formats. For example, grayscale images from DEX bytecode are analyzed with CNNs \cite{paper1}, fused ResNet-CBAM images from bytecode utilize CNN architectures \cite{paper20}, and Markov images from opcodes are processed via CNN+DNN hybrids for family classification \cite{paper21}. Permission-based features extracted from manifest files are commonly used with ML models such as SVM \cite{paper5} or pruned for reduced complexity with minimal loss of discriminative power \cite{paper47}. Opcode sequences and n-grams are applied with algorithms like MMLVS \cite{paper3} or Random Forest \cite{paper57}, while entropy histograms from DEX binaries detect packing \cite{paper17}. API-based features include frequency vectors and structural features analyzed with ensemble models \cite{paper30} or API flows captured through dominance trees for behavioral classification \cite{paper39}. Graph-based representations, including pruned call graphs \cite{paper66} and sensitive function call graphs (SFCG) from APIs \cite{paper8}, leverage Random Forests to capture semantic and relational information. DNNs have also been applied successfully to permission vectors \cite{paper40}.

\subsection{Dynamic analysis Features}

\begin{figure}[htbp]
\centering
\includegraphics[width=\columnwidth]{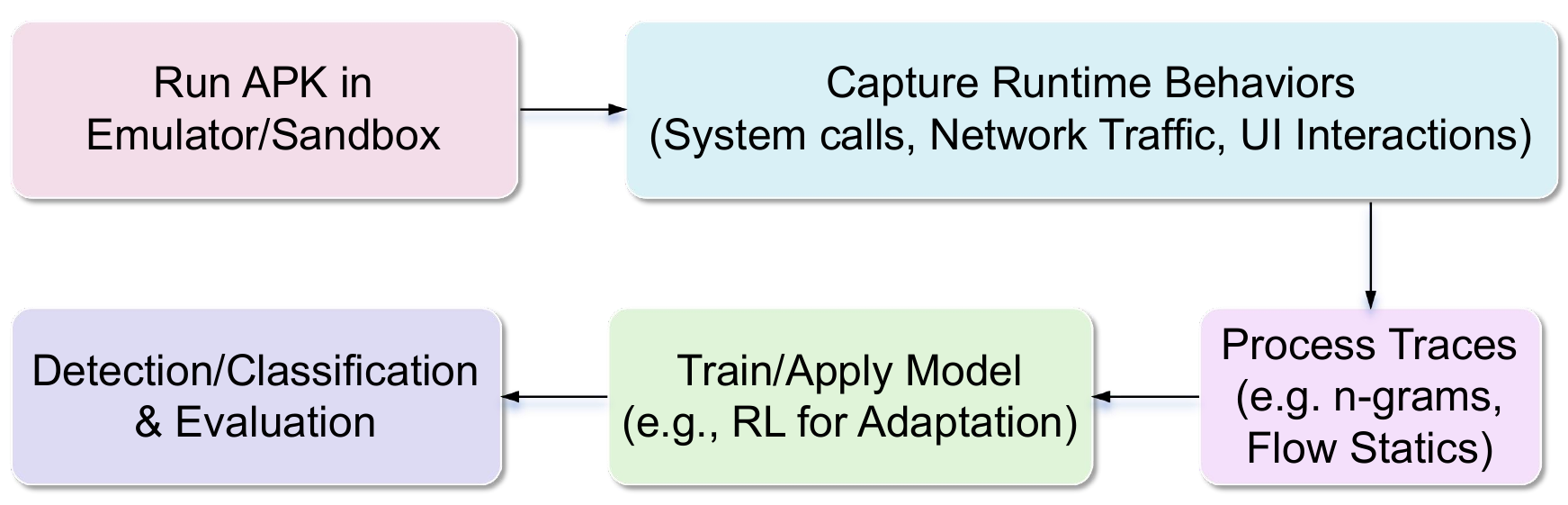}
\caption{Workflow for Static Analysis Technique in Android Malware Detection.}
\label{fig:dynamic-workflow}
\end{figure}

Illustrated in Fig. \ref{fig:dynamic-workflow}, dynamic analysis focuses on runtime observation to uncover behaviors that may remain hidden during static inspection, making it particularly effective against obfuscation and polymorphism in Android malware. Typically, APKs are executed in controlled environments such as emulators, sandboxes, or testbeds \cite{paper4,paper13,paper25,paper38,paper41,paper63,paper67,paper68}, which simulate real usage while avoiding risks to physical devices. During execution, diverse features are captured, including system calls \cite{paper4,paper25,paper49}, network traffic traces \cite{paper12,paper22,paper58,paper67,paper68}, behavioral logs of UI or API interactions \cite{paper41,paper13}, and even concolic execution paths \cite{paper63}. To address noise and scalability issues, these raw features are further processed through techniques such as n-gram on traces, swarm optimization for feature reduction \cite{paper38,paper22}, or damped statistical updates in reinforcement learning (RL) \cite{paper25}. The refined data is then used for model training, where methods like RL are adopted for drift adaptation, metric learning for few-shot scenarios \cite{paper12}, and open set recognition (OSR) for handling unseen threats \cite{paper67,paper68}. Despite its robustness against encrypted and polymorphic malware, dynamic analysis still faces challenges including computational overhead, emulator divergences \cite{paper63}, dependency on user interactions \cite{paper41}, and partial blind spots in encrypted traffic \cite{paper12}.

Building on these foundations, one such work is MalBoT-DRL \cite{10283893}, which applies deep reinforcement learning-assisted DNN to detect IoT malware botnets. In this approach, workloads generated by IoT devices are routed through a switch, and the resulting traffic is analyzed. A traffic handler extracts 23 statistical features per time window using a damped incremental statistics method, enabling fast and resource-efficient feature capture. To reduce bias toward benign traffic, normalization is applied, while an attention-based reward mechanism further highlights suspicious behaviors. The processed data is then fed into the MalBoT-DRL engine, which employs a Double Deep Q-Network (DQN) that integrates Q-learning with deep neural networks. Training involves initialization, sampling, experience replay, and overestimation avoidance, where replay buffers stabilize learning and Double DQN prevents Q-value inflation \cite{10.5555/3016100.3016191}. Through this workflow, MalBoT-DRL dynamically updates its strategy, mitigates model drift, and achieves improved generalization to zero-day botnet attacks.

Similarly, Ali \cite{9865131} followed the same dynamic analysis workflow using a multitask deep learning model. Traffic flows are collected from datasets such as IoT-23 and VARIoT, and execution traces are transformed into three distinct modalities, such as flow-related, flag-related, and packet payload, to provide complementary views of network behavior. During processing, missing values are removed, features are normalized with a Min–Max approach, and class imbalance is addressed through SMOTEENN oversampling. To further refine the data, feature selection is performed at both early and late stages: the former applied before modality splitting and the latter after modality fusion, with recursive XGBoost and SULOV methods used to prune redundant variables. The processed data is used to train a multitask LSTM model, which first distinguishes benign from malicious traffic and then classifies malware type. Training is carried out with stratified five-fold cross-validation and hyperparameter optimization to ensure robustness.

\subsubsection{Comparison Among Different Dynamic Analysis Features}
Table \ref{tab:feature-comparison} summarizes the variety of feature extraction techniques employed in dynamic analysis, where features are drawn from different runtime sources and applied to diverse detection tasks. For instance, system call traces such as ioctl are monitored for family classification \cite{paper4}, while abstracted traces and n-gram representations improve detection efficiency \cite{paper25}. Behavioral divergences in execution traces reveal emulator-based evasions \cite{paper63}, and packet captures (PCAP) combined with particle swarm optimization (PSO) are used for ransomware detection \cite{paper22}. Other studies extract HTTP headers from traffic for classification \cite{paper58}, session images from encrypted traffic with metric learning for few-shot detection \cite{paper12}, and prioritized traffic features for identifying unknown malware \cite{paper38}. User-triggered behaviors spanning API calls, UI activity, file access, and network usage have also been leveraged for variant detection \cite{paper41}, while open set recognition (OSR) techniques analyze HTTP traffic to handle previously unseen malware \cite{paper67,paper68}.


\subsection{Hybrid Analysis Features}

\begin{figure}[htbp]
\centering
\includegraphics[width=\columnwidth]{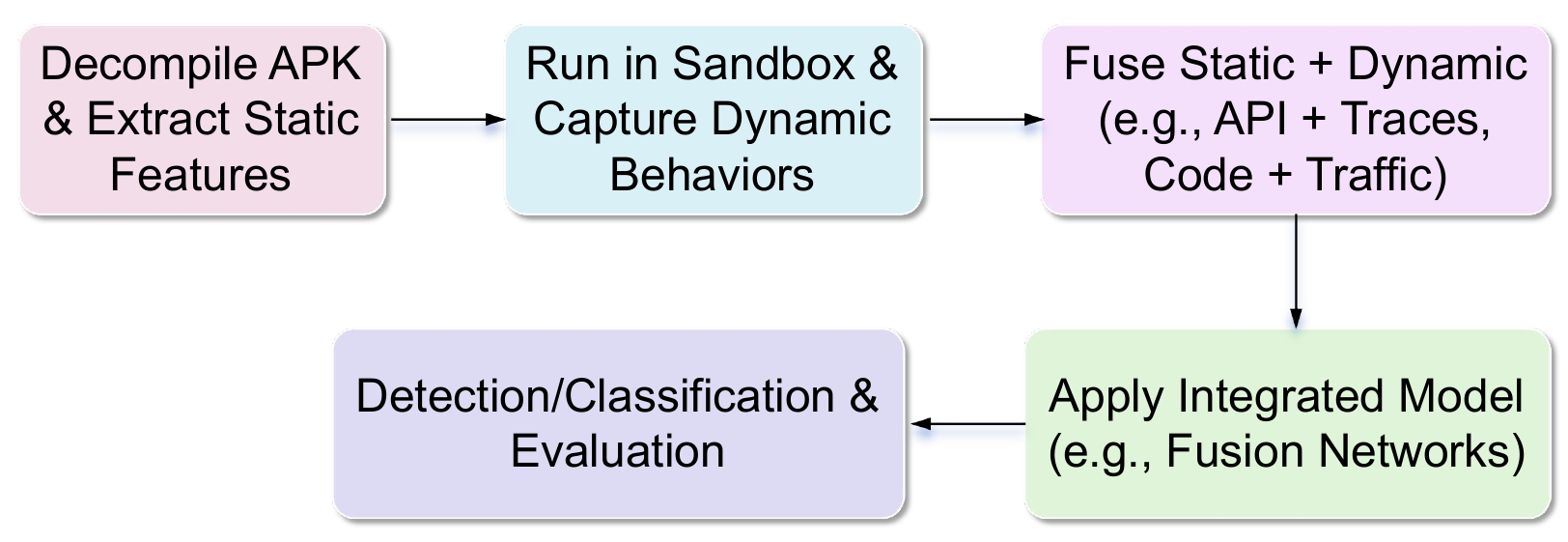}
\caption{Workflow for Static Analysis Technique in Android Malware Detection.}
\label{fig:hybrid-workflow}
\end{figure}

The hybrid workflow in Fig. \ref{fig:hybrid-workflow} merges static decompilation with dynamic execution to achieve comprehensive threat coverage. Static analysis provides efficient baselines by extracting features such as permissions or text for OCR fusion \cite{paper45}, APIs and opcodes for behavioral pattern analysis \cite{paper13}, or control-flow graphs (CFGs) for structural insights \cite{paper61}. Dynamic execution in sandboxes complements this by capturing runtime behaviors, including traces and network traffic for fusion \cite{paper64}, screenshots enabling OCR-based ransomware detection \cite{paper45}, or grammar-based traffic patterns for command-and-control (C2) detection in C2Miner \cite{10.1145/3634737.3644992}. Fusion further enhances coverage, for example by combining code and traffic \cite{paper61} or integrating activation and disambiguation for detecting live servers in C2Miner, often using multi-view or grammar-based methods to mitigate gaps. At the modeling stage, ensembles and deep neural networks are widely adopted \cite{paper13,paper41,paper60}. Overall, hybrid approaches strengthen robustness against zero-day and polymorphic threats, but they also face challenges such as the added complexity of dual-phase pipelines, reliance on feature quality (e.g., OCR precision \cite{paper45}), and assumptions such as the availability of malware binaries, as in C2Miner.

A representative system that follows this workflow is C2Miner, proposed by Davanian \cite{10.1145/3634737.3644992} to detect and analyze live botnet C2 servers. The system introduces three core components: a sandbox to activate malware binaries, a profiler to disambiguate traffic and assess C2 likelihood, and a man-in-the-middle (MitM) module to redirect communications toward candidate servers. The workflow begins with two inputs—an IoT malware binary and a target IP:port space. Once the binary is executed in the sandbox, the profiler filters irrelevant protocols, measures how frequently a target is contacted, and calculates a likelihood score indicating potential C2 activity. The MitM module then redirects traffic: for IP/port-based targets it substitutes addresses directly, while for DNS-based targets it relies on host-file mappings or alternative resolution strategies. To determine whether a target is truly a C2 server, two complementary methods are applied. The SYN-DATA–aware method evaluates communication sequences, though it may misclassify in rare cases where malware repeatedly opens and closes connections. To address this limitation, the fingerprinting-aware method models communication using a grammar-based approach, providing more reliable identification and enabling clustering of C2 communication patterns. Finally, the framework outputs a list of live servers together with tags that can be used to identify C2 traffic in broader network traces.

\subsubsection{Comparison Among Different Hybrid Analysis Features}
Table \ref{tab:feature-comparison} provides a comprehensive comparison of hybrid feature extraction techniques. In \cite{paper13}, static APIs and opcodes are fused with dynamic execution traces, with pattern models applied for malware detection. In \cite{paper64}, static opcode and API features are combined with runtime instrumentation, which are then represented as Markov images and processed through fused neural networks. In \cite{paper45}, static permissions are integrated with dynamic OCR analysis of screenshots, enabling machine learning models to support ransomware detection. In \cite{paper61}, control-flow graphs are combined with network flow features, with classifiers used for traffic categorization. In \cite{paper18}, static and dynamic features are partitioned and then fused to reduce false negatives during classification. Finally, C2Miner \cite{10.1145/3634737.3644992} extracts features by coupling sandbox activation with grammar-based traffic disambiguation, using algorithmic solutions to identify and fingerprint command-and-control servers.


\subsection{Graph Learning Representation Features}

\begin{figure}[htbp]
\centering
\includegraphics[width=\columnwidth]{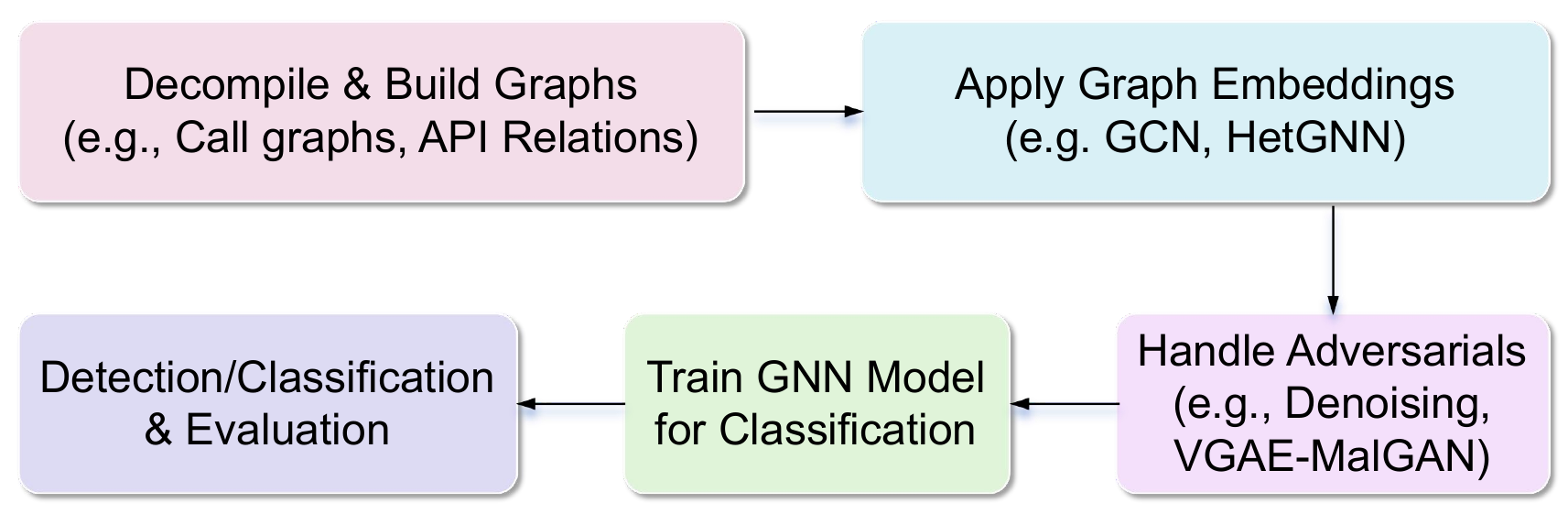}
\caption{Workflow for Static Analysis Technique in Android Malware Detection.}
\label{fig:graph-workflow}
\end{figure}

As shown in Fig. \ref{fig:graph-workflow}, graph-based analysis models the relational structures of applications to achieve semantic-rich detection, offering robustness against adversarial manipulation in Android malware. The workflow begins with graph construction, typically by decompiling applications and extracting elements such as sensitive APIs, opcodes, or control-flow relations. For example, function call graphs (CG) are built from Smali/DEX code \cite{paper8}, control flow graphs (CFG) integrate opcode features \cite{paper50}, semantic data flow chains capture method-level semantics \cite{paper36}, and heterogeneous relations incorporate permissions, intents, or other contextual attributes \cite{paper52,paper53}. Embeddings refine these—GCN for convolution \cite{paper8,paper36}, denoising for perturbations  in \cite{paper48}, HetGNN for multi-types in \cite{paper52}, with pruning for efficiency in \cite{paper66}. Adversarial handling adapts via generation in \cite{paper51} or meta-path defenses in \cite{paper53}. GNN training classifies families (e.g., SVM on embeddings). Strengths encompass deep insights against obfuscation/polymorphic threats, interpretability via subgraphs \cite{paper50,paper66}.

 A representative example is VGAE-MalGAN  \cite{9814995}, which combines graph neural networks with generative adversarial networks to improve robustness against adversarial attacks. The workflow begins with application decompilation, where Smali and Manifest files are extracted using Apktool. From the Smali files, APIs are identified with their code block IDs, and feature selection via linear regression is applied to construct local function call graphs, which are then merged into a global graph. In the next step, The centrality features are selected  with Permission and Intent features from the Manifest file to train a GNN, which determines whether an application is malicious or benign. To address adversarial threats, VGAE-MalGAN introduces a generator–substitute detector architecture. The generator perturbs malware graphs by adding edges or nodes sampled from benign applications, while the substitute detector mimics the target classifier to learn its decision boundaries. This adversarial process continues until the substitute detector fails to distinguish the modified graphs. Finally, the augmented dataset, containing both original and adversarial graphs, is used to retrain the GNN-based classifier, thereby hardening it against future attacks.

\subsubsection{Comparison Among Different Graph Learning Representation Features}

Table \ref{tab:feature-comparison} summarizes the range of graph learning representation features proposed in the literature. In \cite{paper8}, sensitive function call graphs (SFCGs) are constructed from APIs, with graph convolutional networks (GCNs) applied for malware detection. In \cite{paper36}, semantic data flow chains are built to capture method-level semantics, which are then classified using GCNs. In \cite{paper48}, subgraph networks (SGNs) are employed to model local dependencies, and denoising GCNs are used to improve robustness against obfuscation. In \cite{paper50}, call graphs are embedded with opcode vectors and combined with linear SVMs for family identification. In \cite{paper52}, weighted heterogeneous graphs (WHGs) integrate multiple relation types, and HetGNN is applied for malware detection. In \cite{paper53}, adversarial heterogeneous graphs with meta-path representations are designed to capture cross-type relations, with SVMs used for family classification under adversarial conditions. In \cite{paper51}, adversarial API graphs are generated via GAN to simulate attacks, and GNN-based models are retrained for hardened detection. Finally, in \cite{paper66}, pruned call graphs are constructed to reduce graph complexity, with Random Forests applied for behavioral classification.


\begin{table*}[htbp]
\centering
\caption{Comparison of Feature Extraction Techniques in Android Malware Analysis (Performance Ranges from Key Examples like \cite{paper1,paper64,paper52})}
\label{tab:feature-comparison}
\begin{adjustbox}{width=\textwidth}
\begin{tabular}{|p{2cm}|p{2.5cm}|p{3cm}|p{2.5cm}|p{2.5cm}|p{3cm}|}
\hline
\textbf{Analysis Type} & \textbf{Feature} & \textbf{Description} & \textbf{Datasets} & \textbf{Performance} & \textbf{Strengths / Limitations} \\ \hline
\multirow{5}{*}{Static} & Image-based (DEX to images) & Converts bytecode to grayscale/RGB/Markov images, used with CNN/ViT models & Drebin, VirusShare, IMG\_DS & 94–99.5\% accuracy & Robust to noise, high precision; requires preprocessing and compute-intensive \\ \cline{2-6}
 & API calls \& Function Call Graphs (FCG) & Extracts sensitive APIs and structural relations; applied with GCN & Drebin, AndroZoo, MalDroid & 89–98.22\% accuracy & Captures semantic behavior; graph construction overhead \\ \cline{2-6}
 & Permissions & Lightweight feature vectors from manifest permissions & CICMalDroid, Drebin & 90–97.4\% accuracy & Efficient, scalable; vulnerable to obfuscation \\ \cline{2-6}
 & Opcodes \& N-grams & Opcode sequences or Markov images from bytecode & Drebin, AndroZoo & 81–95.7\% accuracy & Detects structural patterns; high dimensionality risk \\ \cline{2-6}
 & Entropy-based & Entropy histograms of DEX binaries & Drebin, VirusShare & $\sim$94\% precision & Good for detecting packing/obfuscation; limited behavioral depth \\ \hline
\multirow{5}{*}{Dynamic} & System Call Tracing & Runtime system/kernel calls, n-grams, abstractions & Drebin, TwinDroid & 94–97.9\% accuracy & Resists obfuscation; emulator detection, resource overhead \\ \cline{2-6}
 & Network Traffic Analysis & PCAP flows, HTTP requests, encrypted traffic & CICAndMal2017, CICInvesAndMal2019 & 80–94\% accuracy & Robust to code obfuscation; limited by encryption \\ \cline{2-6}
 & Behavioral Profiling & Captures UI/API/permission behaviors at runtime & Custom traces, Drebin+benign & $>$96\% variant detection & Comprehensive behaviors; requires user interaction \\ \cline{2-6}
 & Dynamic Symbolic Execution (Concolic) & Hybrid concrete + symbolic execution & Lumus, testbeds & Reported in related works & Explores multiple paths; computationally expensive \\ \cline{2-6}
 & Reinforcement Learning Models & RL-driven traffic analysis (e.g., MalBoT-DRL) & MedBIoT, N-BaIoT & 99.4–99.8\% accuracy & Adapts to drift, zero-day; latency, complexity \\ \hline
\multirow{4}{*}{Hybrid} & Static + Dynamic Fusion & Combines API/opcode with runtime traces & Drebin, CICMalDroid & 97–99.52\% accuracy & Improves robustness; overhead from multi-view processing \\ \cline{2-6}
 & Code + Network Traffic & CFG/structural + traffic features & Drebin, CICAndMal2017 & 97\% detection / 94\% categorization & Captures semantics and runtime; high complexity \\ \cline{2-6}
 & C2 Traffic Detection (C2Miner) & Sandbox activation + grammar-based C2 disambiguation & MalwareBazaar, VirusTotal traces & 92\% precision, 86\% F1 & Detects live C2 servers; assumptions may fail with adaptive malware \\ \cline{2-6}
 & OCR/Text + Static Features & OCR on screenshots/logs combined with permission features & RansomProber, APKPure & $\sim$94\% accuracy & Detects ransomware text; quality dependent \\ \hline
\multirow{5}{*}{Graph Learning} & Sensitive Function Call Graphs (SFCG) + GCN & Graphs of sensitive APIs with semantic/triadic features & Drebin & 98.22\% accuracy & Robust to obfuscation; requires anchor APIs \\ \cline{2-6}
 & Semantic Data Flow Chains + GCN & Control/data flow chains with embeddings & Drebin, custom & 95.8\% accuracy & Captures deep semantics; higher complexity \\ \cline{2-6}
 & Denoising GCN (SGN) & Subgraph networks resilient to adversarial perturbations & CICMalDroid, Drebin & 97.1\% F1 & Robust against adversarial samples; preprocessing overhead \\ \cline{2-6}
 & Weighted Heterogeneous Graphs + HetGNN & Multi-type graph with relation-aware embeddings & AndroZoo, Tencent HG & 97.92\% F1 & Fuses multi-relations; vulnerable without defenses \\ \cline{2-6}
 & VGAE-MalGAN (Adversarial) & GAN generates adversarial API graphs for retraining & Drebin, CICMalDroid & 95–99\% (post-defense) & Strengthens classifiers; assumes retraining availability \\ \hline
\end{tabular}
\end{adjustbox}
\end{table*}

\section{Opportunities and Challenges}
\subsection{Comparison Among Different Analysis Techniques}
The reviewed papers demonstrate that static analysis offers efficiency for on-device deployment, achieving accuracies like 98.12\% in MCADS \cite{paper1} and 99.5\% in EAODroid \cite{paper31}, but is vulnerable to obfuscation, as transformations reduce detection by over 60\% \cite{paper27}. Dynamic analysis resists evasion better, with system call abstraction yielding 97.9\% \cite{paper25} and HTTP-based OSR 92\% F1 at low openness \cite{paper67}, but incurs overhead and emulator detection risks \cite{paper63}. Hybrid methods balance these, fusing opcode/API Markov for 99.52\% \cite{paper64} or code/network for 97\% \cite{paper61}, though complex. Graph learning excels in relational modeling, with GCN that uses function call graphs at 98.22\% \cite{paper8} and HetGNN that uses weighted heterogeneous graphs at 97.92\% \cite{paper52}, robust to attacks (93.5\% F1 post-defense \cite{paper53}) versus static's 73.53\% under CW \cite{paper9}, but requires graph construction.

\begin{figure}[htbp]
\centering
\includegraphics[width=\columnwidth]{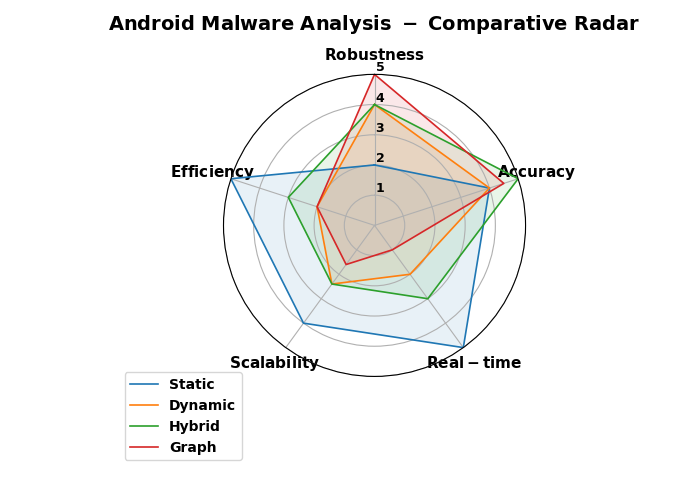}  
\caption{Comparative Radar Chart of Android Malware Analysis Techniques across Key Metrics (Scores 1-5 derived from reviewed papers; higher is better). Static excels in efficiency, dynamic in robustness, hybrid offers balance, and graph leads in accuracy.}
\label{fig:radar-comparison}
\end{figure}
The radar chart in Fig. \ref{fig:radar-comparison} provides a multidimensional comparison of the four primary Android malware analysis techniques—Static, Dynamic, Hybrid, and Graph-based—across five key performance metrics: Robustness, Efficiency, Accuracy, Scalability, and Real-time capability. Derived from averaged insights in the 68 reviewed papers, the chart uses a 1-5 scale (1 being low performance and 5 high), with each axis representing a metric and the colored polygons illustrating the relative strengths of each technique. For instance, the blue Static polygon extends prominently toward Efficiency and Scalability, reflecting its lightweight nature for rapid, resource-efficient processing (e.g., permissions achieving 97.4\% accuracy with minimal features in \cite{paper5}), but contracts on Robustness due to susceptibility to obfuscation attacks (e.g., detection drops $>60\%$ in \cite{paper27}). The orange Dynamic shape peaks at Robustness, capturing runtime resilience against code manipulations (e.g., system call abstractions at 97.9\% in \cite{paper25}), yet dips in Efficiency and Real-time owing to emulation overheads (e.g., resource intensity in \cite{paper63}). Hybrid (green) forms a balanced profile, with strong Accuracy from fused features (e.g., 99.52\% in opcode/API Markov integration \cite{paper64}), mitigating individual weaknesses while maintaining moderate scores across other metrics. Finally, the red Graph-based polygon excels in Accuracy and Robustness through relational modeling (e.g., 97.92\% F1 in heterogeneous graphs \cite{paper52} and 97.1\% under perturbations in \cite{paper48}), but retracts on Real-time and Scalability due to computational demands of graph construction (e.g., in \cite{paper53}). Overall, this visualization underscores the trade-offs highlighted in the SLR: while no technique is superior across all dimensions, hybrid and graph-based methods offer the most promise for comprehensive, resilient detection in evolving Android threat landscapes.

\subsection{Performance Trends Over Time}
\begin{figure}[htbp]
\centering
\includegraphics[width=\columnwidth]{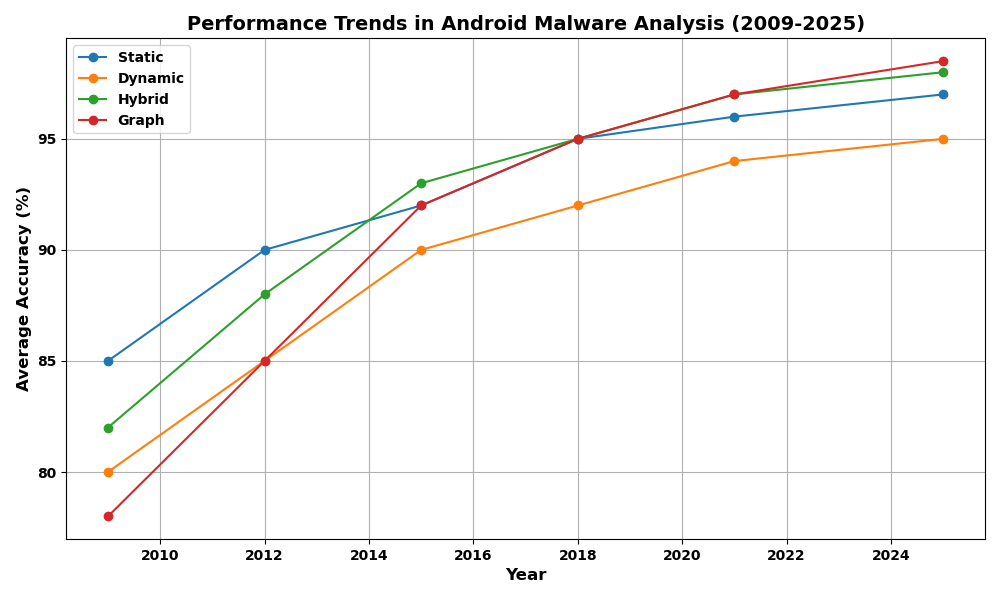}
\caption{Performance Trends in Android Malware Analysis (2009-2025), Showing Average Accuracy Improvements Across Techniques (Derived from Reviewed Papers). Hybrid and Graph methods demonstrate the steepest gains.}
\label{fig:performance-trends}
\end{figure}

Fig. \ref{fig:performance-trends} depicts the evolution of average accuracy in Android malware detection across the four analysis techniques from 2009 to 2025, synthesized from trends in the reviewed papers. The x-axis represents publication years, while the y-axis shows average accuracy percentages, with lines and markers distinguishing Static (blue), Dynamic (orange), Hybrid (green), and Graph-based (red) methods. Data points are approximated averages from reported performances in the summaries, illustrating a general upward trajectory driven by advancements in feature extraction and modeling.
The chart reveals that all techniques have improved over time, reflecting the field's maturation. Static analysis starts at ~85\% (e.g., early permission-based models in foundational works like \cite{paper5}) and reaches ~97\% by 2025, benefiting from efficient features like image conversions (98.12\% in \cite{paper1}) but plateauing due to obfuscation limits \cite{paper27}. Dynamic methods begin lower (~80\%, e.g., basic traffic analysis) but climb to ~95\%, bolstered by refinements like system call abstractions (97.9\% in \cite{paper25}) and HTTP OSR (94.33\% in \cite{paper58}), though constrained by runtime overhead \cite{paper63}. Hybrid approaches show steeper gains, from ~82\% to 98\%, leveraging fusions for superior generalization (e.g., 99.52\% in opcode/API Markov \cite{paper64}; 97\% in code+traffic \cite{paper61}). Graph-based techniques exhibit the most rapid rise post-2015 (~78\% to 98.5\%), fueled by relational embeddings and defenses (e.g., 97.92\% F1 in \cite{paper52}; 97.1\% under attacks in \cite{paper48}).

This visualization underscores the shift toward hybrid and graph methods for addressing complex threats, with recent accuracies surpassing 97\% highlighting opportunities for real-time, robust systems in Android ecosystems.

\subsection{Future Research}
Based on the synthesis of insights from the 68 reviewed papers, several promising avenues emerge for advancing Android malware detection. These directions address key gaps identified in the analysis, such as the need for greater integration across techniques, improved robustness against evolving threats, and enhanced scalability for real-world deployment. Below, I elaborate on the suggested areas, providing rationale, potential approaches, and links to relevant papers for context.

\subsubsection{Integrate Dynamic with Static Features for Hybrid Approaches to Counter Integrated Threats}
Hybrid methods already show strong potential by combining the efficiency of static analysis (e.g., permissions and API calls) with the evasion-resistance of dynamic analysis (e.g., runtime behaviors). However, future work should focus on deeper integration to tackle "integrated threats" like polymorphic malware that exploit both code obfuscation and runtime adaptations. For instance, fusing static features (e.g., opcode patterns from \cite{paper18}) with dynamic user-triggered traces (e.g., API/UI behaviors in \cite{paper41}) or network flows (e.g., in \cite{paper60}) could create adaptive models that detect threats in fragmented Android environments. This could involve automated feature selection algorithms to minimize overhead, building on multiview fusions that achieved 99.52\% accuracy \cite{paper64}. Rationale: Current hybrids often underutilize temporal runtime data, leading to missed zero-day variants; enhanced integration would improve generalization, as seen in partial successes against evasion \cite{paper63}.

\subsubsection{Extend Graph-based Techniques to Inter-procedural Flows or Runtime Data}
Graph learning excels in capturing relational semantics (e.g., API call graphs in \cite{paper50}), but most implementations are limited to intra-procedural or static structures. Future research should extend graphs to inter-procedural flows (e.g., dominance trees for API mining in \cite{paper39}) or incorporate runtime data (e.g., dynamic embeddings in heterogeneous graphs from \cite{paper52}). This could involve hybrid graph models that fuse static FCGs with dynamic traces, enabling detection of runtime evasions while maintaining robustness (97.92\% F1 in \cite{paper52}). Rationale: Polymorphic malware often hides in inter-method calls or runtime behaviors; addressing scalability issues in large graphs is needed (e.g., 96.1\% reduction via pruning in \cite{paper66}), making them viable for on-device use. 

\subsubsection{Enhance Adversarial Robustness via Probability-based Risks or Integrated Defenses}
Adversarial attacks remain a critical vulnerability, with drops like 51.68\% F1 in poisoned graphs \cite{paper53}. Future efforts should prioritize probability-based risk assessment (e.g., federated learning privacy in \cite{paper32}) or integrated defenses like denoising (97.1\% F1 under perturbations in \cite{paper48}) and retraining (98.68\% post-defense in \cite{paper51}). This could include ensemble models combining GCN with attention mechanisms for meta-path walks \cite{paper53}. Rationale: As malware evolves with AI-generated evasions, probabilistic defenses would quantify uncertainty, improving resilience beyond current baselines (e.g., 73.53\% drop under CW attacks in \cite{paper9}).

\subsubsection{Explore Real-time Deployment and Cross-platform Applications}
Real-time detection is underrepresented, with few on-device hybrids (e.g., patterns in \cite{paper13} or chains in \cite{paper35}). Future work should optimize for Android's resource constraints, such as lightweight models for lifespan labeling \cite{paper37} or edge computing in TransMalDE. Cross-platform extensions (e.g., adapting to iOS/Windows in \cite{paper33}) could generalize frameworks. Rationale: Android's fragmentation demands real-time scalability; cross-platform would address multi-OS threats, building on traffic prioritization reducing features by 59\% \cite{paper38}.

\section{CONCLUSION AND DISCUSSION}

This systematic literature review (SLR) has provided a comprehensive examination of feature extraction techniques in Android malware analysis, synthesizing insights from 68 papers published between 2009 and 2025. By categorizing approaches into static, dynamic, hybrid, and graph-based paradigms, we have elucidated their effectiveness in addressing the escalating threats of sophisticated malware. Static features, like permissions and opcodes, offer efficiency and scalability for on-device deployment but remain susceptible to obfuscation. Dynamic methods, including system call tracing and network traffic analysis, enhance robustness against code manipulations, albeit at the cost of resource intensity. Hybrid techniques balance these by fusing static and dynamic elements, achieving superior accuracy and generalization. Graph-based representations excel in capturing relational semantics, demonstrating high resilience to adversarial attacks, though with computational overheads.

While this SLR advances understanding of Android malware detection, several limitations warrant discussion. First, the selection criteria (relevance-based filtering) may exclude shorter, innovative works or non-English publications, potentially biasing toward established methods. Second, reliance on reported metrics (e.g., accuracy/F1) assumes consistency across studies, but variations in datasets and evaluation setups (e.g., imbalanced classes in Drebin) limit direct comparisons. Third, the focus on Android-specific threats overlooks emerging cross-platform risks, though papers like \cite{paper33} hint at this need.
Broader implications include the need for standardized benchmarks with temporal metadata to combat model drift, as seen in outdated datasets like Genome. Practically, our roadmap—emphasizing hybrid-graph integrations and adversarial defenses—can guide developers toward more secure Android ecosystems. Future SLRs could incorporate real-time tools or meta-analyses for quantitative synthesis.
In conclusion, as Android malware evolves, advancing feature extraction through integrated, robust techniques will be crucial for safeguarding user privacy and system integrity.



%
{\footnotesize \bibliographystyle{unsrt}}
\bibliography{Transactions-Bibliography/example}

\end{document}